\documentclass[12pt]{article}
\usepackage[centertags]{amsmath}
\usepackage{amsfonts}
\usepackage{amssymb}
\usepackage{amsthm}
\usepackage{graphicx}
\usepackage{wrapfig}

\newcommand{\beq}{\begin{equation}}
\newcommand{\eeq}[1]{\label{#1}\end{equation}}
\newcommand{\bea}{\begin{eqnarray}}
\newcommand{\eea}[1]{\label{#1}\end{eqnarray}}

\def\a{\alpha}
\def\b{\beta}

\def\G{\Gamma}
\def\d{\delta}
\def\D{\Delta}
\def\e{\epsilon}

\def\h{\eta}

\def\l{\lambda}
\def\L{\Lambda}
\def\m{\mu}
\def\n{\nu}

\def\p{\pi}

\def\r{\rho}

\def\s{\sigma}

\def\t{\tau}

\def\vf{\varphi}
\def\ps{\psi}
\def\Ps{\Psi}

\def\nb{\nabla}

\begin{document}
\setlength{\topmargin}{-1cm} \setlength{\oddsidemargin}{0cm}
\setlength{\evensidemargin}{0cm}

\begin{titlepage}
\begin{center}
{\Large \bf Higher-Spin Modes in a Domain-Wall Universe}

\vspace{20pt}

{\large Manuela Kulaxizi and Rakibur Rahman}

\vspace{12pt}

Universit\'e Libre de Bruxelles \& International Solvay Institutes\\
ULB-Campus Plaine C.P. 231, B-1050 Bruxelles, Belgium

\end{center}
\vspace{20pt}

\begin{abstract}
We find a consistent set of equations of motion and constraints for massive higher-spin fluctuations in a gravitational
background, required of certain characteristic properties but more general than constant curvature space. Of particular
interest among such geometries is a thick domain wall$-$a smooth version of the Randall-Sundrum metric. Apart from the
graviton zero mode, the brane accommodates quasi-bound massive states of higher spin contingent on the bulk mass.
We estimate the mass and lifetime of these higher-spin resonances, which may appear as metastable dark matter in a
braneworld universe.
\end{abstract}

\end{titlepage}

\newpage

\section{Introduction}\label{sec:Intro}

Consistent interacting theories of higher-spin (HS) fields are difficult to construct. For massless fields, interactions are generically
in tension with HS gauge invariance, and such pathologies lead to various no-go theorems in flat space~\cite{Old,gpv,Aragone,ww,New}.
Even the free propagation in non-trivial backgrounds may suffer from difficulties. Noticed long ago by Fierz and Pauli~\cite{FP}, the
latter kind of problem shows up for massive fields at the level of equations of motion (EoMs) and constraints by rendering them
mutually incompatible. A Lagrangian formulation takes care of this issue, but the resulting system is likely to propagate unphysical
modes or allow propagation outside the light cone~\cite{vz,sham,kob,d3}. Appropriate non-minimal terms may come to the rescue
and provide a consistent Lagrangian description of free massive HS fields in backgrounds with constant
curvature~\cite{AN2,PRS,Buchbinder:2012iz,Buch,Met}.

Is it possible to describe consistently the free propagation of a massive field of arbitrary spin in spaces more general than the
constant curvature ones? The answer is yes, at least at the level of EoMs and constraints, as we will show in this paper. The
necessary conditions require only that the following irreducible Lorentz tensors\footnote{The
notation $(i_1\cdots i_n)$ means totally symmetric expression in all the indices $i_1,\dots,i_n$ with the normalization factor
$\tfrac{1}{n!}$. The totally antisymmetric expression $[i_1\cdots i_n]$ comes with the same normalization.}, characteristic of the
$D$-dimensional background metric $g_{\m\n}$, vanish:
\bea X_{\m\n\r}{}^{\a\b}&\equiv&\nb_{(\m}W_\n{}^\a{}_{\r)}{}^\b-\left(\tfrac{2}{D+2}\right)g_{(\m\n}\nb^\s W_{\r)}{}^{(\a}
{}_\s{}^{\b)}=0,\label{Xdefined}\\Y_{\m\n\r}&\equiv&\nb_{(\m}R_{\n\r)}-\left(\tfrac{2}{D+2}\right)g_{(\m\n}\nb_{\r)}R=0,
\label{Ydefined}\\Z_{\m\n\r}&\equiv&2\nb_{[\r}R_{\m]\n}+\left(\tfrac{1}{D-1}\right)g_{\n[\r}\nb_{\m]}R+\left(\m
\leftrightarrow\n\right)=0,\eea{Zdefined} where $W_{\m\n\r\s}$ is the Weyl tensor, $R_{\m\n}$ the Ricci tensor, and $R$
the scalar curvature. In such a geometry, the consistent set of dynamical equations and constraints describing a probe totally
symmetric spin-$s$ bosonic field $\vf_{\m_1...\m_s}$ will be given by
\beq \left[\nb^2-M^2+\tfrac{2(s-1)(s+D-2)}{(D-1)(D+2)}\hat{R}\right]\vf_{\m_1...\m_s}+s(s-1)\hat{R}_{(\m_1}{}^{\r}{}_{\m_2}{}^{\s}
\vf_{\m_3\dots\m_s)\r\s}-s\hat{R}_{\r(\m_1}\vf^{\r}{}_{\m_2\dots\m_s)}=0,\eeq{KG-grav}\vspace{-30pt}
\bea &\nb\cdot\vf_{\m_1\dots\m_{s-1}}\equiv\nb^{\m_s}\vf_{\m_1\dots\m_s}=0,&\label{Div-grav}\\&\vf^\prime_{\m_1\dots\m_{s-2}}
\equiv g^{\m_{s-1}\m_s}\vf_{\m_1\dots\m_s}=0,&\eea{Tr-grav}
where the quantity $\hat{R}_{\m\n\r\s}$ is the Riemann tensor minus its constant trace part,
\beq \hat{R}_{\m\n\r\s}\equiv R_{\m\n\r\s}-\tfrac{2\L}{(D-1)(D-2)}\left(g_{\m\r}g_{\n\s}-g_{\n\r}g_{\m\s}\right),\eeq{R-hat}
that conveniently parametrizes the deviation of the manifold under consideration from a constant curvature space of cosmological
constant $\L$, and $M$ is the mass in the latter. The assumptions include locality and that neither any vacuum expectation values,
which possibly source the geometry, nor any other fluctuations show up at the linearized level. The existence of an underlying
Lagrangian formulation, however, is not assumed.

A number of interesting geometries satisfy the conditions~(\ref{Xdefined})--(\ref{Zdefined}). Symmetric spaces have covariantly
constant Riemann tensors: $\nabla_\l R_{\m\n\r\s}=0$, and therefore qualify. Some coset spaces arising from supergravity and
M-theory compactifications
as well as some pp-wave backgrounds are of this kind. In particular, the well-known
$\text{AdS}_5\times\text{S}^5$ geometry of string theory, or in fact any $\text{AdS}_p\times\text{S}^q$ even with unequal radii,
is a symmetric space.

We will see that certain domain-wall (DW) geometries of phenomenological interest also fulfill the
conditions~(\ref{Xdefined})--(\ref{Zdefined}). DW spacetimes in general arise naturally from a system of gravity plus
scalar(s) with a potential. They play an important role in describing holographic renormalization group flows. Because
there is an FLRW cosmology corresponding to every DW solution of a given model~\cite{Skenderis:2006fb}, these geometries are
also interesting in the context of inflationary cosmology. Moreover, the Randall-Sundrum one-brane model~\cite{RS} may find
smooth generalizations through some DW solutions~\cite{DeWolfe} (see also Refs.~\cite{Thick} and references therein, for example).
Among the DW geometries that satisfy the conditions~(\ref{Xdefined})--(\ref{Zdefined}), there is indeed one that serves as
a thick-brane realisation of the braneworld. The HS fluctuations on this geometry, governed by the
Eqs.~(\ref{KG-grav})--(\ref{Tr-grav}), may therefore have phenomenologically interesting consequences.

The organization of this paper is as follows. The next section, which the reader may skip without loss of continuity, employs
the ``involutive deformation method'' to derive the consistent set of EoMs and
constraints~(\ref{KG-grav})--(\ref{Tr-grav}) describing the free propagation of a massive spin-$s$ field in a gravitational
background subject to the conditions~(\ref{Xdefined})--(\ref{Zdefined}). Some technical details of this section are relegated
to Appendix~\ref{sec:ID}. In Section~\ref{sec:DW}, we show that certain DW metrics with maximally symmetric slicings do
fulfill the aforementioned criteria. In particular, there exists a smooth generalization of the Randall-Sundrum metric that
also qualifies. We briefly recall the consequences the fluctuations of the latter geometry bring along, i.e., a localized
graviton zero mode and a continuum of Kaluza-Klein modes on the thick brane. Section~\ref{sec:HSflc} considers HS fluctuations
on top of this background. As the transverse traceless modes of the highest-spin field on the brane decouple completely from
any other mode, the equivalent Schr\"odinger problem for them can be easily studied. Thankfully, normalizable HS zero modes
are ruled out, but massive HS resonances on the brane are allowed. The mass and lifetime of these metastable HS states are
estimated. We make some concluding remarks in Section~\ref{sec:Conclusions}, notably that these HS resonances in a domain-wall
universe may be so long lived as to qualify as dark matter candidates without contradicting the tests of the inverse-square law
of gravity.

\section{Massive HS Fields in a Gravitational Background}\label{sec:Background}

A massive spin-$s$ bosonic field in flat space is customarily represented by a rank-$s$ symmetric traceless Lorentz
tensor, say $\vf_{\m_1\dots\m_s}$. It satisfies the dynamical Klein-Gordon equation:
\beq I_{\m_1\dots\m_s}\equiv\left(\partial^2-\mathfrak m^2\right)\vf_{\m_1\dots\m_s}=0,\eeq{KG}
and is subject to the divergence and trace constraints:
\bea &J_{\m_1\dots\m_{s-1}}\equiv\partial\cdot\vf_{\m_1\dots\m_{s-1}}=0,&\label{Div}\\
&K_{\m_1\dots\m_{s-2}}\equiv\vf'_{\m_1\dots\m_{s-2}}=0.&\eea{Tr}
The divergence and trace constraints are crucial in the counting of propagating degrees of freedom $\mathfrak{D}$.
In $D$ spacetime dimensions, it is given by
\beq \mathfrak{D}=2\,{D-4+s\choose s-1}+{D-4+s\choose s},\eeq{DoF}
which of course reduces to $2s+1$ in $D=4$.

On the other hand, the mutual compatibility of the dynamical equation and constraints is indispensable for a consistent description.
In other words, Eqs.~(\ref{KG})--(\ref{Tr}) can be viewed as an involutive system of differential equations~\cite{Inv-book}, that fulfill
the ``gauge identities'':
\bea &\mathcal G_{1,\,\m_1\dots\m_{s-1}}\equiv\partial\cdot I_{\m_1\dots\m_{s-1}}-\left(\partial^2-\mathfrak m^2\right)
J_{\m_1\dots\m_{s-1}}=0,&\label{GIi}\\&{\mathcal G}_{2,\,\m_1\dots\m_{s-2}}\equiv I'_{\m_1\dots\m_{s-2}}
-\left(\partial^2-\mathfrak m^2\right)K_{\m_1\dots\m_{s-2}}=0,&\label{GIii}\\&{\mathcal G}_{3,\,\m_1\dots\m_{s-3}}
\equiv J'_{\m_1\dots\m_{s-3}}-\partial\cdot K_{\m_1\dots\m_{s-3}}=0.&\eea{GIiii}
thanks to the commutativity of ordinary derivatives. The above gauge identities however are not all independent, since the
trace of $\mathcal G_{1,\,\m_1\dots\m_{s-1}}$ can be expressed in terms of $\mathcal G_{2,\,\m_1\dots\m_{s-2}}$ and
$\mathcal G_{3,\,\m_1\dots\m_{s-3}}$. In other words, there is a gauge identity for the gauge identities:
\beq \mathcal H_{\m_1\dots\m_{s-3}}\equiv\mathcal G'_{1,\,\m_1\dots\m_{s-3}}-\partial\cdot\mathcal G_{2,\,\m_1\dots\m_{s-3}}
+\left(\partial^2-\mathfrak m^2\right)\mathcal G_{3,\,\m_1\dots\m_{s-3}}=0.\eeq{GIGI}
From the point of view of an involutive system, the mutual compatibility of Eqs.~(\ref{KG})--(\ref{Tr}) is taken care of by the
gauge identities~\cite{Involution}. It was shown long ago~\cite{Boss} that the degrees of freedom count is related to 
the ``strength of the system''. An explicit expression for $\mathfrak D$ is given in Ref.~\cite{Involution} in terms of the number 
of equations $t_k$ and independent gauge identities $l_k$ of order $k$ in derivatives:
\beq \mathfrak D=\tfrac{1}{2}\sum_k k(t_k-l_k).\eeq{dof}
Indeed, this formula reproduces the count~(\ref{DoF}) with the correct values of $t_k$ and $l_k$:
\bea t_k&=&\d_k^2\,{D+s-1\choose s}+\d_k^1\,{D+s-2\choose s-1}+\d_k^0\,{D+s-3\choose s-2},\label{NEq}\\[5pt]
l_k&=&\d_k^3\,\left[{D+s-2\choose s-1}-{D+s-4\choose s-3}\right]+\d_k^2\,{D+s-3\choose s-2}+\d_k^1\,{D+s-4\choose s-3}.\eea{NGI}

Consistency requires that any deformation of the flat-space free system~(\ref{KG})--(\ref{Tr}) always fulfills the gauge identities.
However, in a gravitational background, for example, the na\"ive covariantization $\partial_\mu\rightarrow\nb_\mu$ of the
flat-space system results in algebraic inconsistencies, since covariant derivatives no longer commute. Noticed already in
Ref.~\cite{FP}, such problems are in fact very generic for HS systems. For some special backgrounds, though, they may be cured
by the addition of non-minimal terms. An explicit example of this appears below.

To consider the free propagation of a massive spin-$s$ particle in a gravitational background, we first deform the
system~(\ref{KG})--(\ref{Tr}) into the following:
\bea &I_{\m_1\dots\m_s}\equiv\left(\nb^2-\mathfrak m^2\right)\vf_{\m_1\dots\m_s}+\D I_{\m_1\dots\m_s}=0,&\label{KG1}\\
&J_{\m_1\dots\m_{s-1}}\equiv\nb\cdot\vf_{\m_1\dots\m_{s-1}}+\D J_{\m_1\dots\m_{s-1}}=0,&\label{Div1}\\
&K_{\m_1\dots\m_{s-2}}\equiv\vf'_{\m_1\dots\m_{s-2}}+\D K_{\m_1\dots\m_{s-2}}=0.&\eea{Tr1}
where the non-minimal deformations $\D I_{\m_1\dots\m_s}$, $\D J_{\m_1\dots\m_{s-1}}$ and $\D K_{\m_1\dots\m_{s-2}}$ are linear in the
field $\vf_{\m_1\dots\m_s}$, and contain at least one power of the curvature. They only contain lower-derivatives of the field lest
unphysical modes should appear or causal propagation be lost. The involutive deformation method~\cite{Involution} consists of
finding the deformations~(\ref{KG1})--(\ref{Tr1}), for which there exists a deformed version of the relations~(\ref{GIi})--(\ref{GIiii}),
i.e.,
\beq \mathcal G_{i,\,\a_1\dots\a_{s-i}}\equiv{\mathcal I}^{~\,\m_1\dots\m_s}_{i,\,\a_1\dots\a_{s-i}}\,I_{\m_1\dots\m_s}
+{\mathcal J}^{~\,\m_1\dots\m_{s-1}}_{i,\,\a_1\dots\a_{s-i}}\,J_{\m_1\dots\m_{s-1}}+{\mathcal K}^{~\,\m_1\dots\m_{s-2}}
_{i,\,\a_1\dots\a_{s-i}}\,K_{\m_1\dots\m_{s-2}}=0,\eeq{GI1}
where the operators $\mathcal I_i$, $\mathcal J_i$, $\mathcal K_i$ with $i=1,2,3$ are called the gauge identity generators. Again,
they are minimal deformations of the free theory plus non-minimal corrections:
\beq {\mathcal I}^{~\,\m_1\dots\m_s}_{i,\,\a_1\dots\a_{s-i}}~=~\d_i^1\,\d^{(\m_1\dots\m_{s-1}}_{\a_1\dots\a_{s-1}}\nb^{\m_s)}~+~\d_i^2\,
\d^{(\m_1\dots\m_{s-2}}_{\a_1\dots\a_{s-2}}g^{\m_{s-1}\m_s)}~+~\D {\mathcal I}^{~\,\m_1\dots\m_s}_{i,\,\a_1\dots\a_{s-i}}\,,\eeq{op1}
\beq{\mathcal J}^{~\,\m_1\dots\m_{s-1}}_{i,\,\a_1\dots\a_{s-i}}~=~\d_i^1\,\d^{\m_1\dots\m_{s-1}}_{\a_1\dots\a_{s-1}}
\left(-\nb^2+\mathfrak m^2\right)~+~\d_i^3\,\d^{(\m_1\dots\m_{s-3}}_{\a_1\dots\a_{s-3}}g^{\m_{s-2}\m_{s-1})}~+~\D {\mathcal J}
^{~\,\m_1\dots\m_{s-1}}_{i,\,\a_1\dots\a_{s-i}}\,,\eeq{op2}\vspace{-10pt}
\beq{\mathcal K}^{~\,\m_1\dots\m_{s-2}}_{i,\,\a_1\dots\a_{s-i}}~=~
\d_i^2\,\d^{\m_1\dots\m_{s-2}}_{\a_1\dots\a_{s-2}}\left(-\nb^2+\mathfrak m^2\right)~-~\d_i^3\,\d^{(\m_1\dots\m_{s-3}}_{\a_1\dots\a_{s-3}}
\nb^{\m_{s-2})}~+~\D {\mathcal K}^{~\,\m_1\dots\m_{s-2}}_{i,\,\a_1\dots\a_{s-i}}\,.\eeq{op3}

In Appendix~\ref{sec:ID}, we have shown how the gauge identities~(\ref{GI1}) may be satisfied under the assumption of locality. It turns out
the first gauge identity, $\mathcal G_{1,\,\a_1\dots\a_{s-1}}=0$, can be fulfilled, with a free parameter $\a$, modulo that we set to zero
certain anomalous terms containing derivatives of the curvature. These bad terms are given in Eq.~(\ref{badGR}), and in order for them to
vanish it is necessary that the gravitational background satisfy the conditions~(\ref{Xdefined})--(\ref{Zdefined}) for generic spin,
namely $X_{\m\n\r}{}^{\a\b}=0$, $Y_{\m\n\r}=0$ and $Z_{\m\n\r}=0$. The vanishing of the last term in Eq.~(\ref{badGR}) further requires:
\beq \nb_\m R=0,\qquad \text{or}\qquad \a=\frac{2(s-1)(s+D-2)}{(D-1)(D+2)}\,.\eeq{alpha}
Now the freedom of the parameter $\a$ plays a crucial role. By choosing $\a$ to the above value, one may be able to do with
a background of non-constant Ricci scalar: $\nb_\m R\neq0$.

Under these conditions all the gauge identities can be fulfilled, with non-minimal corrections to the equations and gauge identity
generators given by Eqs.~(\ref{0dKG1})--(\ref{0dGI13}) and Eqs.~(\ref{dKG1})--(\ref{dGI2}). These corrections in principle
contain $\mathcal O(R^2)$-terms. However, they do not contribute at $\mathcal O(R^2)$, but only at $\mathcal O(R^3)$, in the gauge 
identities:
\beq \D{\mathcal I}^{~\,\m_1\dots\m_s}_{i,\,\a_1\dots\a_{s-i}}\D I_{\m_1\dots\m_s}
+\D{\mathcal J}^{~\,\m_1\dots\m_{s-1}}_{i,\,\a_1\dots\a_{s-i}}\D J_{\m_1\dots\m_{s-1}}
+\D{\mathcal K}^{~\,\m_1\dots\m_{s-2}}_{i,\,\a_1\dots\a_{s-i}}\D K_{\m_1\dots\m_{s-2}}=\mathcal O(R^3).\eeq{GI12}
This means that in the deformations~(\ref{KG1})--(\ref{Tr1}) all the the higher-curvature terms can be consistently set to zero.
The resulting system has undeformed divergence and trace:
\bea &\left[\nb^2-\mathfrak m^2+\a R\right]\vf_{\m_1...\m_s}+s(s-1)R_{(\m_1}{}^{\r}{}_{\m_2}{}^{\s}
\vf_{\m_3\dots\m_s)\r\s}-s R_{\r(\m_1}\vf^{\r}{}_{\m_2\dots\m_s)}=0,&\label{00KG-grav}\\
&\nb\cdot\vf_{\m_1\dots\m_{s-1}}=0,&\label{00Div-grav}\\&\vf^\prime_{\m_1\dots\m_{s-2}}=0.&\eea{00Tr-grav}
This system is consistent, under the conditions~(\ref{Xdefined})--(\ref{Zdefined}) and~(\ref{alpha}), up to all orders in the
curvature. Note that the addition of $\mathcal O(R^2)$ terms, which is inessential for consistency, may require further
conditions. For a background with a non-constant Ricci scalar, $\a$ must be set to the value of Eq.~(\ref{alpha}). The EoMs and
constraints~(\ref{KG-grav})--(\ref{Tr-grav}) then follow from incorporating the constant trace part of the curvature tensor into
the mass term.

One still needs to check that there exists a deformed counterpart of the identity~(\ref{GIGI}). A straightforward computation gives
\bea &&\mathcal G'_{1,\,\m_1\dots\m_{s-3}}-\nb\cdot\mathcal G_{2,\,\m_1\dots\m_{s-3}}+\left(\nb^2-\mathfrak m^2\right)
\mathcal G_{3,\,\m_1\dots\m_{s-3}}\nonumber\\[5pt]&=&-(s-3)(s-4)R_{(\m_1}{}^{\r}{}_{\m_2}{}^{\s}\mathcal G_{3,\,\m_3\dots\m_{s-3})\r\s}
+(s-3)R^\r{}_{(\m_1}\mathcal G_{3,\,\m_2\dots\m_{s-3})\r}-\a R\,\mathcal G_{3,\,\m_1\dots\m_{s-3}}\nonumber\\[5pt]&&
+\left[\,\a-\tfrac{2(s-3)(s+D-4)}{(D-1)(D+2)}\,\right]\left(\nb^\r R\right)K_{\m_1\dots\m_{s-3}\,\r}
+\D\mathcal K^{\prime\,\a_1\dots\a_{s-2}}_{1,\,\m_1\dots\m_{s-3}}K_{\a_1\dots\a_{s-2}}\,.\eea{GIGI1}
Therefore, $\mathcal G'_{1,\,\m_1\dots\m_{s-3}}$ can be expressed completely in terms of $\mathcal G_{2,\,\m_1\dots\m_{s-2}}$ and
$\mathcal G_{3,\,\m_1\dots\m_{s-3}}$ provided that the last line in Eq.~(\ref{GIGI1}) vanishes. For $\nb_\m R=0$, the latter
condition is automatic. When $\nb_\m R\neq0$, there are two possibilities: one is to start with a field whose trace is vanishing
identically rather than just as an on-shell condition~\cite{Involution,Nacho}. In this case, $K_{\m_1\dots\m_{s-2}}$ would never appear
in the system and its reduced number of gauge identities. By so doing, one would demand that the trace always remain zero, even in the
presence of interactions\footnote{Such a requirement may have non-trivial consequences in a possible Lagrangian formulation of the system.
We thank I.~L.~Buchbinder and Y.~M.~Zinoviev for stressing out this point.}. Another possibility is to view our original
system~(\ref{KG})--(\ref{Tr}) as the zero-trace gauge fixing of a system of symmetric rank-$s$ field with a Weyl symmetry:
$\d\vf_{\m_1\dots\m_s}=g_{(\m_1\m_2}\lambda_{\m_3\dots\m_s)}$. Now the freedom of the rank-$(s-2)$ parameter $\l_{\m_1\dots\m_{s-2}}$
allows one to choose the trace to vanish even at the interaction level\footnote{We are thankful to M.~Taronna for bringing this
possibility to our attention.}. The massless counterpart of such a system is well known in the literature as Conformal Higher
Spin~\cite{Fradkin:1985am,Segal} (See also Refs.~\cite{Tseytlin,CHS,Taeke} for recent discussions).

\section{The Thick Domain Wall}\label{sec:DW}

Let us consider the following domain wall metric in $D=d+1$ dimensions
\beq ds^2=dy^2+e^{2f(y)}\left[-(1-k r^2)dt^2+\frac{dr^2}{1-k r^2}+r^2 d\Omega_{d-2}\right],\qquad -\infty<y<+\infty,\eeq{dwm}
where $k=(-1,0,+1)$ correspond respectively to $d$-dimensional AdS, flat and dS slicings. We would like to see if such a geometry
can possibly satisfy the conditions~(\ref{Xdefined})--(\ref{Zdefined}). Because the metric~(\ref{dwm}) is conformally flat,
$X_{\m\n\r}{}^{\a\b}=0$ automatically. It turns out that $Z_{\m\n\r}=0$ for any $f(y)$ as well. The only non-trivial condition on
the metric is imposed by the vanishing of the tensor $Y_{\m\n\r}$; it requires $f(y)$ to satisfy the following differential equation:
\beq f'''-2f'f''-4k e^{-2f}f'=0.\eeq{f-cond}
The generic solution of this equation is given in terms of Jacobi elliptic functions:
\beq f(y)=-\ln{\left[a\,\text{sn}\left(b+\frac{y}{l},\,ka^2l^2\right)\right]}.\eeq{gen}
where $a$, $b$ and $l\neq0$ are constants.

For $k=0$, the metric~(\ref{dwm}) boils down to one with $d$-dimensional Poincar\'e invariance:
\beq ds^2=dy^2+e^{2f(y)}\h^{ij}dx_idx_j,\eeq{DW-RS0}
where $\h^{ij}$ is the flat metric with $i,j=0,1,...,d-1$. The general solution~(\ref{gen}), on the other hand,
reduces to
\beq f(y)=-\ln{\left[a\cosh{\left(b+\frac{y}{l}\right)}\right]},\eeq{fours}
which of course obeys the differential equation
\beq f''=f^{\prime\,2}-l^{-2}.\eeq{reducedDE}
The solution~(\ref{fours}) satisfies the null energy condition with real $b$ and $l$, since
\beq T_t^t-T_y^y=-\,\frac{3}{l^2}\,\text{sech}^2\left(b+\frac{y}{l}\right)\leq0.\eeq{null-DW}

Note that pure $\text{AdS}_{d+1}$ of radius $l=\sqrt{-d(d-1)/2\L}$\, solves Eq.~(\ref{reducedDE}) with $f'=l^{-1}$ and $f''=0$.
For more generic solutions~(\ref{fours}), therefore, the quantity $f''$ will parametrize the deviation from AdS space.
In other words, the ``hatted'' Riemann tensor $\hat{R}_{\m\n\r\s}$ defined in Eq.~(\ref{R-hat}) will be proportional to $f''$.
Indeed, its non-zero content is given by
\beq \hat{R}^i_j=-(d+1)\d^i_j f'',\qquad \hat{R}^y_y=-2d f'',\qquad \hat R=-d(d+3)f''.\eeq{hatted-objects}

The thick-brane solution we will be interested in corresponds to a simple choice of parameters: $a=1$ and $b=0$
in Eqs.~(\ref{DW-RS0})--(\ref{fours}). This gives
\beq f(y)=-\ln\cosh\left(\frac{y}{l}\right),\eeq{DW-RS}
which represents a smooth generalization of the Randall-Sundrum metric~\cite{RS}, the thickness of the brane being $\mathcal O(l)$.
This particular thick-brane generalization has already been studied in Ref.~\cite{Gremm}. Note that the metric~(\ref{DW-RS}) is
conformally flat with a non-constant Ricci scalar, and does not asymptote to AdS space.

We will consider massive HS fluctuations in this geometry. Although, in the context of braneworlds, the massless case
has been studied by some authors~\cite{Germani,Ferrara}, no study of the massive ones seems to be present. But first let us
discuss briefly the graviton fluctuations.

\subsubsection*{Graviton Fluctuations}\label{sec:graviton}

Universal aspects of graviton fluctuations in conformally flat backgrounds preserving $d$-dimensional Poincar\'e invariance have
been extensively studied in the literature. Here we follow Ref.~\cite{Csaki}. From the $d$-dimensional point of view, graviton
fluctuations of the form $h_{ij}(x,y)=\ps(y)\,\e_{ij}\,e^{iq\cdot x}$ will obey the following equation in the transverse traceless gauge:
\beq \left[\,\partial_y^2+(d-4)f'\partial_y-e^{-2f}q^2-2f''-2(d-2)f^{\prime 2}\,\right]\ps(y)=0,\eeq{graviton0}
which can be derived from the Einstein equations in the bulk. Note that in the next section we are going to present a
generalization~(\ref{TT-KG}) of this equation for the transverse traceless modes of a fluctuation of arbitrary spin and mass.
For a massless graviton in the bulk, with $M^2l^2=-2$~\cite{Metsaev}, indeed the general equation reduces to the above one.

The existence of normalizable $d$-dimensional modes is connected with the asymptotic behavior of the potential of the equivalent
Schr\"odinger problem.
It turns out there are no normalizable negative energy graviton modes (with $-q^2<0$).
For $-q^2=0$, there is a normalizable mode~\cite{Gremm} given by
\beq \ps_0(y)=\sqrt{\frac{3}{4 l}}\,\text{sech}^2\left(\frac{y}{l}\right),\eeq{graviton1}
which is identified as the localized massless graviton on the brane.

There are no massive graviton bound states nor any resonances~\cite{Gremm}, but a continuum of Kaluza-Klein modes for all $-q^2>0$, as
they usually appear~\cite{RS,Csaki}. This can be shown, for example, from the generalized case of the next section.
These Kaluza-Klein modes will alter the behavior of gravity at length scale $\mathcal O(l)$~\cite{RS}. In particular, Newton's
inverse square law will get modified, and this poses an upper bound on $l$ from table-top experiments~\cite{Kapner}. The bound
turns out to be $l\lesssim10^{-4}\,\text{m}$.

\section{Higher-Spin Fluctuations}\label{sec:HSflc}

In conformally flat backgrounds, in general, the HS dynamical equation~(\ref{KG-grav}) reduces to
\beq \left[\nb^2-M^2+\b\hat{R}\right]\vf_{\m_1\dots\m_s}-\tfrac{s(2s+d-3)}{d-1}\,\hat{R}^\r{}_{(\m_1}\vf_{\m_2\dots\m_s)\r}
+\tfrac{s(s-1)}{d-1}\,\hat{R}^{\r\s}g_{(\m_1\m_2}\vf_{\m_3\dots\m_s)\r\s}=0,\eeq{hseom}
where $\b=\tfrac{(s-1)\left[s(3d+1)+2(d-1)^2\right]}{d(d-1)(d+3)}$. Along with the divergence and trace constraints, this
equation is suitable for describing small fluctuations of HS fields in the domain-wall geometries listed above.
Let us consider higher-spin fluctuations of the form:
\beq \vf_{\m_1\dots\m_s}(x,y)=\int\frac{d^dq}{(2\p)^d}\,\tilde{\vf}_{\m_1\dots\m_s}(q,y)\,e^{iq\cdot x}.\eeq{HS-Fluc}
on the flat DW background~(\ref{DW-RS0}). The component of $\tilde\vf_{\m_1\dots\m_s}(q,y)$ with $r$ indices in the $y$-direction
($0\leq r\leq s$) will appear as the Fourier transform of a spin-($s-r$) field to an observer on the brane.
The transverse traceless modes of the spin-$s$ field ($r=0$) decouple completely from the other fields at the level
of EoMs and constraints; they satisfy
\beq \left[\,\partial_y^2+(d-2s)f'\partial_y-e^{-2f}q^2-M^2+s(s-d-1)-2(d-1)(s-1)f''\,\right]\tilde\vf(y)=0,\eeq{TT-KG}
where we have suppressed the indices and $q$-dependence of $\tilde\vf$.
For $f(y)$ given by Eq.~(\ref{DW-RS}), the above equation can be brought into the Schr\"odinger form through the following redefinitions
of coordinate and variable:
\beq u=\sinh y,\qquad \Psi(u)=\left[1+u^2\right]^{2s-d+1\over4}\tilde\vf(u),\eeq{redef}
where we have set $l=1$ for simplicity. Thus one arrives at
\beq \left[-\partial_u^2+V(u)\right]\Psi(u)=-q^2\Psi(u),\eeq{schr}
where the potential $V(u)$ is of the form
\beq V(u)=\frac{A u^2}{(1+u^2)^2}+\frac{M^2-B}{1+u^2}\,,\eeq{potential}
with the coefficients $A$ and $B$ depending on the spin and dimensionality as follows:
\bea A&=&\tfrac{1}{4}\left(2s+d-3\right)^2-1,\label{potential1}\\
B&=&A-s-\tfrac{1}{4}\left(d^2-1\right)<A.\eea{potential2}
Note that for all $s\geq1$ and $d\geq3$ we have $A\geq0$, and $M^2\geq B-A$. The latter fact follows from the generalization of
the BF bound~\cite{BFbound} on the AdS mass for $s\geq 1$~\cite{Metsaev}:
%
\beq M^2\,\geq\,s^2+s(d-5)-2(d-2).\eeq{BFHS0}

\begin{figure}
\centering
    \includegraphics[width=.7\textwidth]{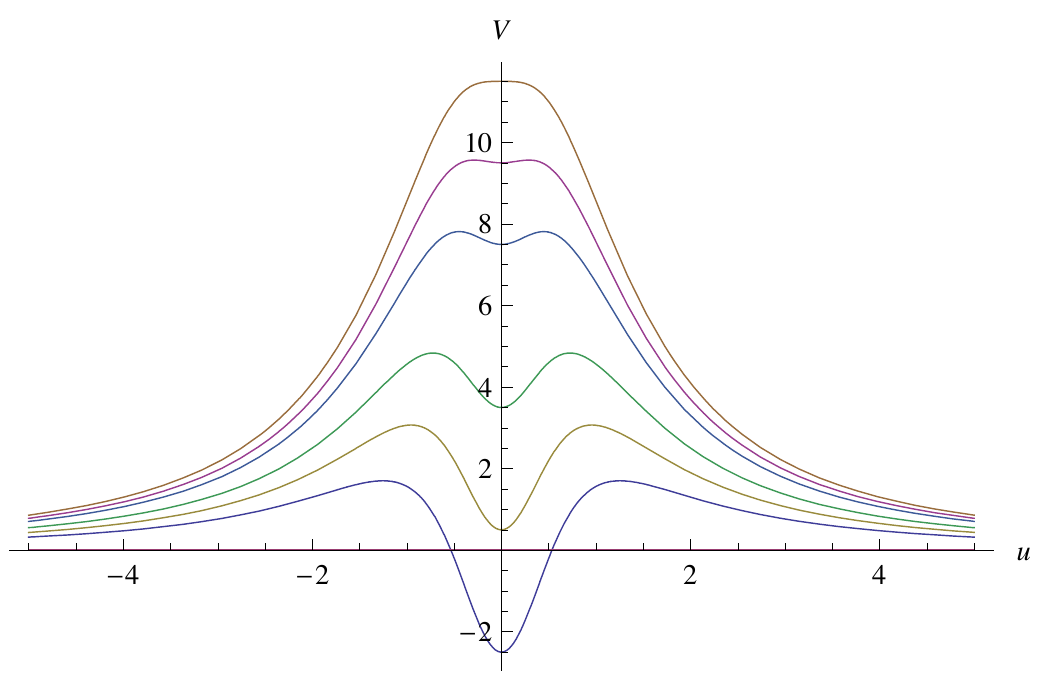}
  \caption{The potential $V(u)$ for a particle of spin $s=3$ in $d=4$ dimensions and values of the bulk mass $M^2=2,5,8,12,14,16$. }
\end{figure}

The potential $V(u)$ is symmetric under reflection, $u\rightarrow-u$, and vanishes as $u\rightarrow\pm\infty$. It has a distinct volcano
shape for the following range of the bulk mass:
\beq B-A<M^2<B+A\,.\eeq{m-star}
The local minimum appears at $u=0$, and the two maxima at $u=\pm\sqrt{\tfrac{A+B-M^2}{A-B+M^2}}$\,. The crater goes above zero at $M^2=B$.
The minimum and maxima disappear for $M^2\geq B+A$, in which case a bell-shaped potential shows up (see Fig.1).

To study the spectrum on the domain wall, let us first note that $q^2$ is to be interpreted as the momentum squared of the $d$-dimensional
fields. As a consistency check one needs to ensure that normalizable tachyonic modes do not exist. Indeed, it is easy to see that
Eq.~(\ref{schr}) does not admit non-trivial solutions for $-q^2<0$ that vanishes at infinity. Below we discuss the (im)possibility of having
localized massless and massive HS modes.

\subsubsection*{Zero Modes and Absence Thereof}

Massless modes correspond to $-q^2=0$, for which the solution of Eq.~(\ref{schr}) is given in terms of associated Legendre polynomials
for generic values of the parameters:
\beq \Psi(u)=\sqrt{1+u^2}\left[\,c_1\,P_\nu^\mu(iu)+c_2\,Q_\nu^\mu(iu)\,\right],\eeq{zero-mode}
where
\beq \n=\sqrt{M^2+\left(\tfrac{d}{2}\right)^2+s}-\tfrac{1}{2},\qquad \m=s+\tfrac{1}{2}(d-3).\eeq{mu-nu}
There exist no normalizable solutions for generic $\m,\n$. But when $\n=\m-n-1$, with $n\in \mathbb N$, the associated Legendre
polynomials do not constitute a set of independent solutions: one solution is of hypergeometric type, while the other is given
by $\Psi(u)= (1+u^2)^{1-\m\over 2} p_n(u)$, where $p_n(u)$ is a polynomial of degree $n$, which is
even(odd) for even(odd) $n$. In this case, the asymptotic behavior of the wave function is $\Ps(u)\sim u^{-\n}$. For $s\geq1$ and
$d\geq3$, both $\n$ and $\m$ are positive, and normalizable higher-spin zero modes seem to show up.

Upon inclusion of the coupling to dynamical gravity in the bulk, this would suggest the existence of gravitationally coupled massless
higher-spin fields on the flat domain wall. This is however in direct contradiction with old~\cite{Old,Aragone} and new~\cite{New}
no-go theorems, which can actually be combined to completely rule out any gravitational coupling of massless higher spins in
flat space~\cite{Combine}. The resolution of the puzzle lies in the values of the bulk mass yielding the zero modes.
The relation between $\n$ and $\m$ gives:
\beq M^2=(n-s+2)(n-s-d+2)-s,\qquad n=0,1,\dots,\tfrac{1}{2}(2s+d-4).\eeq{mass-zero}
But these are precisely the points where the field is (partially) massless in $\text{AdS}_{d+1}$~\cite{DeserPM}.
The points $n\geq 1$ are excluded simply because they fall outside the unitarity region~(\ref{BFHS0}). Neither is the value $n=0$
allowed. To see this, let us note the system~(\ref{KG-grav})--(\ref{Tr-grav}) can be viewed as a deformation around AdS. Now $n=0$
corresponds to a massless field in AdS. However, the associated gauge invariance will be lost in the more generic manifold under
consideration. In other words, the massless case has to be excluded from the beginning for non-constant curvature spaces.
Thus there are no contradictions with the no-go theorems. In the model~\cite{RS}, an apparent contradiction of the similar
kind was seen to arise~\cite{Ferrara}.

\subsubsection*{Massive Quasi-Bound States}

Let us now consider massive modes. When the bulk mass lies within the region~(\ref{m-star}),
the potential acquires a volcano shape and quasi-bound states/resonances may show up\footnote{Bound states are excluded because
for $-q^2\neq0$ the wave function becomes oscillatory as $u\rightarrow\pm\infty$.}.
For an analytic study of the quasi-bound states, let us first rescale the coordinate as:
\beq z\equiv\sqrt 2\left(A+B-M^2\right)^{\frac{1}{4}}u.\eeq{rescale}
Then a Taylor expansion of the potential (\ref{potential}) around $z=0$ reduces Eq.~(\ref{schr}) to the anharmonic
oscillator problem:
\beq \left[-\partial_z^2+\tfrac{1}{4}z^2+\sum_{p=2}^{\infty}\frac{\left(pA+B-M^2\right)z^{2p}}
{\left(-2\sqrt{A+B-M^2}\,\right)^{p+1}}\,\right]\Ps(z)=E\,\Ps(z),\eeq{bender}
where the energy $E$ is related to $-q^2$ as follows:
\beq E=\frac{-q^2-M^2+B}{2\sqrt{A+B-M^2}}\,.\eeq{energy}
For $A\gg1$ and $M^2$ not very close to the upper bound $A+B$, the anharmonic terms can be treated as perturbation. As an approximation,
we will reduce the problem to that of Ref.~\cite{Bender} by keeping only the first term, $\tfrac{1}{4}\lambda\,z^4$, where the perturbation
parameter is
\beq \lambda=-\frac{\left(2A+B-M^2\right)}{2\left(A+B-M^2\right)^{3/2}}<0.\eeq{lambda}
The associated boundary condition $\lim_{|z|\to\infty}\Ps(z)=0$ will select a discrete set of
energy eigenvalues, which are complex~\cite{Bender}. They correspond to metastable states for $|\l|\ll1$. The approximation
of our original problem to that of Ref.~\cite{Bender} will make sense if we restrict ourselves to such eigenfunctions as are
peaked at $z=0$, and have a much lower amplitude away from the origin. Therefore, we will consider only the ground
state of the anharmonic oscillator, for which the energy is given by~\cite{Bender}:
\beq Re(E)\,\approx\,\frac{1}{2}-\frac{3|\l|}{4}\,,\qquad Im(E)\,\approx\,-\,\sqrt{\frac{8}{\pi|\l|}}\,\exp\left(-\frac{1}{3|\l|}
\right).\eeq{E-given}
Note that $Im(E)$ is exponentially small. In view of Eq.~(\ref{energy}), $-q^2$ will also be complex:
\beq-q^2=\left(m-\tfrac{i}{2}\G\right)^2,\eeq{width}
where $m$ is the mass and $\G$ is the width of the metastable state, with $\G\ll m$. Comparing Eqs.~(\ref{energy}), (\ref{E-given})
and~(\ref{width}), one finds that the mass is given by
\beq m^2\,\approx\, M^2-B+\sqrt{A+B-M^2}-\frac{3}{4}\left(\frac{2A+B-M^2}{A+B-M^2}\right),\eeq{anharmonic}
while the lifetime, $\t=1/\G$, is
\beq\t\,\approx\,\left[\frac{\pi\left(2A+B-M^2\right)\left(M^2-B+\sqrt{A+B-M^2}\right)}{32\left(A+B-M^2\right)^{5/2}}\right]^{1/2}
\exp\left[\frac{2\left(A+B-M^2\right)^{3/2}}{3\left(2A+B-M^2\right)}\right].\eeq{lifetime}

\begin{figure}
\centering
    \includegraphics[width=.7\textwidth]{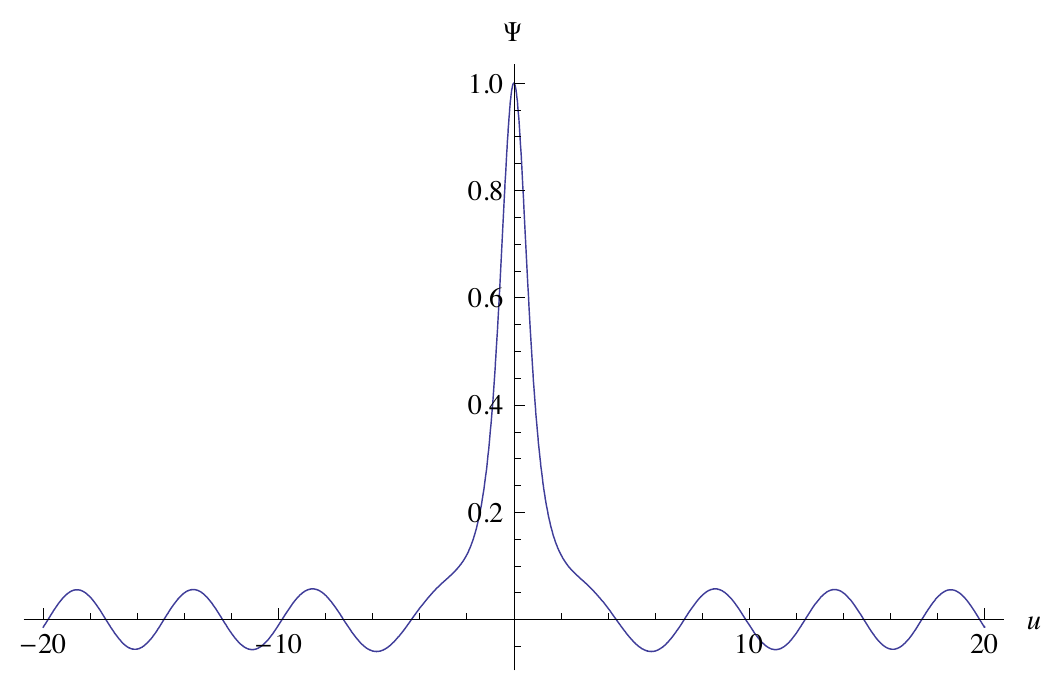}
  \caption{A spin-4 resonance at $m^2=1.687$ in $d=4$, for bulk mass $M^2=10$. The amplitude of oscillations at large $u$ is approximately $0.06$
  with the wave function normalized to unity at the center: $\Psi(u=0)=1$.}
\end{figure}

One may resort to numerics to see if highly-peaked resonances are indeed present. The Schr\"odinger equation~(\ref{schr}),
with the boundary conditions $\Psi(u=0)=1$ and $\Psi'(u=0)=0$, can be solved numerically. The amplitude at $u=0$ is chosen to be unity.
We then scan the solutions for different $m^2$ until we find a solution for which the amplitude of oscillations at infinity is much smaller
than unity. Given a value of the bulk mass in the range~(\ref{m-star}), this procedure gives a single resonant mode at $m^2=m^2(M^2)$
for each $s\geq2$ in $d=4$.

The numerical result for the mass matches well with the value~(\ref{anharmonic}), and therefore to the
ground state energy eigenvalue~(\ref{E-given}). Figs.~2 and 3 show the resonant wave function $\Psi(u)$ for specific values of the bulk mass
and spin in $d=4$ dimensions. For excited-state eigenvalues of the anharmonic oscillator, the wave function around $u=0$ oscillates with an
amplitude comparable to that outside the volcano, and thus the existence of a resonance cannot be established.

\begin{figure}
\centering
    \includegraphics[width=.7\textwidth]{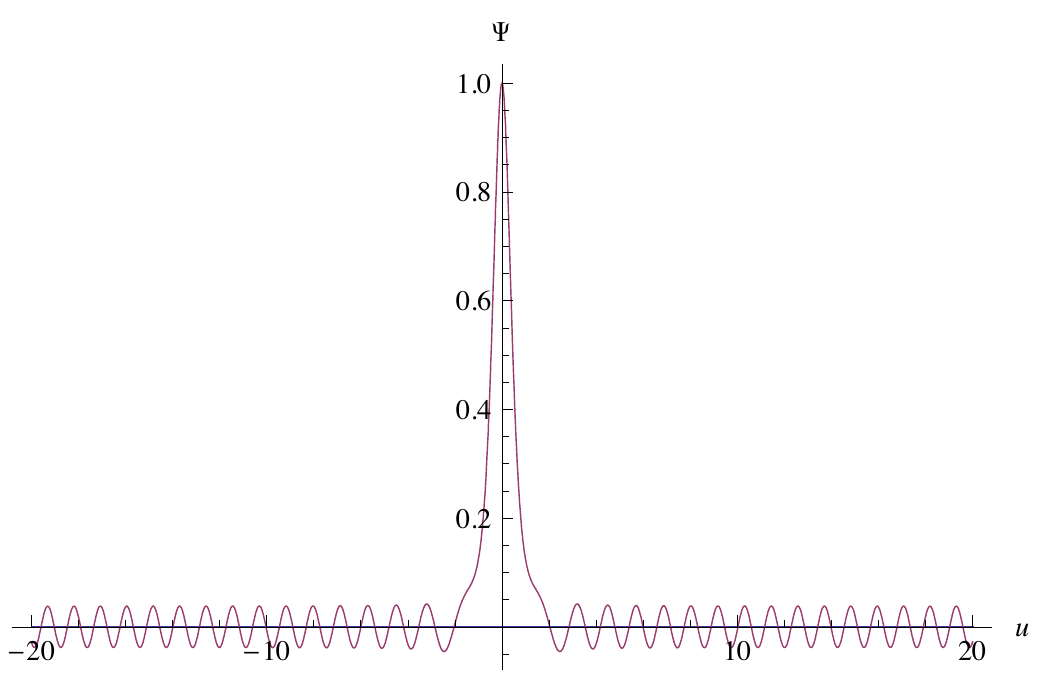}
  \caption{A spin-10 resonance at $m^2=31.900$ in $d=4$, for bulk mass $M^2=120$. The amplitude of oscillations at large $u$ is approximately $0.04$
  with the wave function normalized to unity at the center: $\Psi(u=0)=1$.}
\end{figure}

\section{Concluding Remarks}\label{sec:Conclusions}

In  this paper, we have written down  a consistent set of EoMs and constraints for a free massive HS field propagating in a gravitational
background. The required characteristics of the geometry\footnote{Curiously, the consistency of the Lagrangian dynamics of spinning particles
in various dimensions imposes similar restrictions on the backgrounds~\cite{Spinning,Andrew,Bastianelli}.} allow for spaces of non-constant
curvature. In particular, we found a thick-brane realization of the Randall-Sundrum braneworld that admits consistent free propagation of
massive HS fluctuations. The brane is seen to accommodate not only the graviton but also massive higher-spin resonances, whose mass and
lifetime are estimated.

May these HS modes appear as dark matter in a braneworld universe? The idea of higher-spin dark matter has been explored in Ref.~\cite{ Asorey}.
It is natural for massive HS particles not to couple directly to the Standard Model, and so they are appealing as realistic dark matter candidates.
To qualify as stable dark matter, their lifetime has to exceed the age of the universe: $\tau\gtrsim10^{10}\,\text{years}\sim10^{26}\,\text{m}$.
To see if this is possible in our setup, let us choose for simplicity the typical value $M^2=B$ of the bulk mass. One can reintroduce the
parameter $l$ to rewrite Eqs.~(\ref{anharmonic}) and~(\ref{lifetime}) as
\beq m^2l^2\,\approx\,\sqrt A-\frac{3}{2}\,,\qquad \frac{\t}{l}\approx\sqrt{\frac{\p}{16A}}\,\exp\left(\frac{\sqrt{A}}{3}\right).
\eeq{mass-life}
As already mentioned in Section~\ref{sec:DW}, tests of gravity set $l\lesssim10^{-4}\,\text{m}$~\cite{Kapner}. This means
$\frac{\t}{l}\gtrsim10^{30}$, which corresponds to a relatively stable dark matter particle with spin $s\gtrsim230$.
The mass turns out to be interesting from a phenomenological point is view: $m\gtrsim1\,\text{TeV}$.

We expect these HS particles to couple to gravity like ordinary matter, i.e., to obey the principle of equivalence. In principle, one can go beyond
the free-propagation level and consider gravitational coupling of the HS fields in the bulk. Because the fields are massive, their interactions with
gravity do not suffer from any immediate issues originating from gauge invariance, unlike the massless~\cite{Old,gpv,Aragone,ww,New} and partially massless~\cite{Deser:2012qg,Joung:2014aba,Deser:2013uy} cases.
This is however beyond the scope of our present work. Their interpretation as dark matter necessarily calls for such a study, though. This will be very
important in understanding the details of such dark matter candidates and their possible role in the cosmological evolution of our universe.

Our paper was the first step in trying to describe the propagation of HS fields in DW backgrounds. For simplicity, we did not consider their
coupling to the profile of the scalar field(s) that may source the geometry. It is possible that the inclusion of the scalar profile allow
for more geometries of phenomenological interest. Another interesting direction to pursue is the case of non-zero $k$, i.e., (A)dS slicings.
This may admit some asymptotically AdS geometries that could be studied holographically. We leave this as future work.

\subsection*{Acknowledgments}

We would like to thank R.~Argurio, L.~Calibbi, A.~Campoleoni, A.~Parnachev, M.~Porrati, S.~Rychkov, E.~D.~Skvortsov,
M.~Taronna and A.~Waldron  for useful discussions.
The work of MK was supported in part by the ERC Advanced Grant ``SyDuGraM'', by IISN-Belgium (convention 4.4514.08) and by the
``Communaut\'e Fran\c{c}aise de Belgique" through the ARC program. She gratefully acknowledges the National Science Foundation
Grant No.~PHYS-1066293 and the hospitality of the Aspen Center for Physics, as well as support from the Simons Center for Geometry and
Physics, Stony Brook University during the 2014 Simons Summer Workshop at which some of the research for this paper was performed.
MK is also thankful to the Orthodox Academy of Crete at Kolymbari for hospitality during the ``Quantum Field Theory, String Theory
and Condensed Matter Physics" meeting.
RR is a Postdoctoral Fellow of the Fonds de la Recherche Scientifique-FNRS. His work is partially supported by IISN-Belgium
(conventions 4.4511.06 and 4.4514.08) and by the ``Communaut\'e Fran\c{c}aise de Belgique" through the ARC program. He is grateful to
the Centre of Theoretical Physics at Tomsk State Pedagogical University for its kind hospitality during the conference QFTG'14,
where a part of this work was done.

\begin{appendix}
\numberwithin{equation}{section}

\section{Some Details of the Involutive Deformation}\label{sec:ID}

Our convention for the covariant derivative is: $[\nb_\m,\nb_\n]V^\r=R^\r{}_{\s\m\n}V^\s$. One can write down the various contributions
to the quantity $\mathcal G_{i,\,\a_1\dots\a_{s-i}}$, defined in Eq.~(\ref{GI1}), as:
\bea
\mathcal G_{i,\,\a_1\dots\a_{s-i}}&=&{\D\mathcal I}^{~\,\m_1\dots\m_s}_{i,\,\a_1\dots\a_{s-i}}\,I_{\m_1\dots\m_s}+{\D\mathcal J}^
{~\,\m_1\dots\m_{s-1}}_{i,\,\a_1\dots\a_{s-i}}\,J_{\m_1\dots\m_{s-1}}+{\D\mathcal K}^{~\,\m_1\dots\m_{s-2}}_{i,\,\a_1\dots\a_{s-i}}
\,K_{\m_1\dots\m_{s-2}}\nonumber\\&&+\,\d_i^1\,\left[\,-\mathcal A_{\a_1\dots\a_{s-1}}+\nb\cdot\D I_{\a_1\dots\a_{s-1}}-\left(\nb^2
-\mathfrak m^2\right)\D J_{\a_1\dots\a_{s-1}}\,\right]\nonumber\\&&+\,\d_i^2\,\left[\,\D I'_{\a_1\dots\a_{s-2}}-\left(\nb^2-\mathfrak m^2
\right)\D K_{\a_1\dots\a_{s-2}}\,\right]\nonumber\\&&+\,\d_i^3\,\left[\,\D J'_{\a_1\dots\a_{s-3}}-\nb\cdot\D K_{\a_1\dots\a_{s-3}}\,\right],
\eea{GI2}
where $\mathcal A_{\a_1\dots\a_{s-1}}$ is the sole contribution from the minimal theory:
\beq \mathcal A_{\a_1\dots\a_{s-1}}\equiv\left[\nb^2,\nb^\m\right]\vf_{\m\a_1\dots\a_{s-1}}\neq0,\eeq{A-defined}
which calls for non-minimal corrections to the system~(\ref{KG1})--(\ref{Tr1}) under consideration. This is a 1-derivative term linear
in the curvature. One can use Leibniz rule to extract out of it various other pieces present in the correct gauge identity~(\ref{GI2}),
up to some anomalous terms. Locality admits a unique result up to one free parameter $\a$:
\bea \mathcal A_{\a_1\dots\a_{s-1}}&=&\nb\cdot\D\tilde{I}_{\a_1\dots\a_{s-1}}+{\D\tilde{\mathcal J}}^{~\,\m_1\dots\m_{s-1}}
_{1,\,\a_1\dots\a_{s-1}}\nb\cdot\vf_{\m_1\dots\m_{s-1}}+{\D\tilde{\mathcal K}}^{~\,\m_1\dots\m_{s-2}}_{1,\,\a_1\dots\a_{s-1}}
\vf'_{\m_1\dots\m_{s-2}}\nonumber\\&&+\mathcal B_{\a_1\dots\a_{s-1}}\,,\eea{GR30}
where the first-order correction $\D\tilde I_{\a_1\dots\a_s}$ to the dynamical equation is given by
\beq \D\tilde I_{\a_1\dots\a_s}=s(s-1)R_{(\a_1}{}^{\r}{}_{\a_2}{}^{\s}\vf_{\a_3\dots\a_s)\r\s}-sR_{\r(\a_1}\vf^{\r}{}_{\a_2\dots\a_s)}
+\a R\,\vf_{\a_1...\a_s},\eeq{0dKG1}
and those to the gauge identity generators are
\bea \D\tilde{\mathcal J}^{~\,\m_1\dots\m_{s-1}}_{1,\,\a_1\dots\a_{s-1}}&=&-(s-1)\left[(s-2)\,\d^{\m_1\dots\m_{s-1}}
_{\r\s(\a_1\dots\a_{s-3}}R^\r{}_{\a_{s-2}}{}^\s{}_{\a_{s-1})}-\d^{\m_1\dots\m_{s-1}}_{\r(\a_1\dots\a_{s-2}}
R^\r{}_{\a_{s-1})}\right]\nonumber\\&&-\a R\,\d^{\m_1\dots\m_{s-1}}_{\a_1\dots\a_{s-1}}\,,\label{0dGI12}\\[6pt]
\D\tilde{\mathcal K}^{~\,\m_1\dots\m_{s-2}}_{1,\,\a_1\dots\a_{s-1}}&=&-\tfrac{(s-1)(s-2)}{D-2}\left[\,Y_{(\a_1\a_2}{}^{(\m_1}
\d^{\m_2\dots\m_{s-2})}_{\a_3\dots\a_{s-1})}+\tfrac{4D-7}{3D+6}\,Z_{(\a_1\a_2}{}^{(\m_1}\d^{\m_2\dots\m_{s-2})}_{\a_3\dots
\a_{s-1})}\,\right]\nonumber\\&&-\tfrac{2(s-1)}{D-1}\left[\,\tfrac{s-1}{D+2}\,\nb_{(\a_1}R\,\d^{\m_1\dots\m_{s-1}}_{\a_2\dots\a_{s-1})}
+\tfrac{s-2}{D-2}\,g_{(\a_1\a_2}\nb^{(\m_1} R\,\d^{\m_2\dots\m_{s-2})}_{\a_3\dots\a_{s-1})}\,\right],\eea{0dGI13}
while the remaining anomalous terms $\mathcal B_{\a_1\dots\a_{s-1}}$ read
\bea \mathcal B_{\a_1\dots\a_{s-1}}&=&-\tfrac{(s-1)(s-2)}{D-2}\left[(D-2)X^{\m\n\r}{}_{(\a_1\a_2}\,\vf_{\a_3\dots
\a_{s-1})\m\n\r}+Y^{\m\n\r}\,g_{(\a_1\a_2}\vf_{\a_3\dots\a_{s-1})\m\n\r}\right]\nonumber\\[4pt]&&+\left(\tfrac{s-1}
{D-2}\right)\left[(2s+D-6)\,Y^{\m\n}{}_{(\a_1}\vf_{\a_2\dots\a_{s-1})\m\n}-\left(\tfrac{s+2D-6}{3}\right)
Z^{\m\n}{}_{(\a_1}\vf_{\a_2\dots\a_{s-1})\m\n}\right]\nonumber\\[5pt]&&+\left[\,\tfrac{2(s-1)(s+D-2)}{(D-1)(D+2)}-\a\,\right]
\left(\nb^\m R\right)\vf_{\a_1\dots\a_{s-1}\,\m}\,\eea{badGR}
where $X_{\m\n\r}{}^{\a\b}$, $Y_{\m\n\r}$ and $Z_{\m\n\r}$ are the irreducible Lorentz tensors defined in
Eqs.~(\ref{Xdefined})--(\ref{Zdefined}). These problematic terms vanish if the gravitational background satisfy, for generic spin,
the conditions~(\ref{Xdefined})--(\ref{Zdefined}) plus the condition~(\ref{alpha}).
The first gauge identity, $\mathcal G_{1,\,\a_1\dots\a_{s-1}}=0$, is then fulfilled with the corrected equations:
\bea &\D I_{\m_1\dots\m_s}=\D\tilde{I}_{\m_1\dots\m_s}+\mathcal O(R^2),&\label{dKG1}\\
&\D J_{\m_1\dots\m_{s-1}}=\mathcal O(R^2),&\label{dDiv1}\\
&\D K_{\m_1\dots\m_{s-2}}=\mathcal O(R^2),&\eea{dTr1}
and the corrected gauge identity generators:
\bea &\D\mathcal I^{~\,\m_1\dots\m_s}_{1,\,\a_1\dots\a_{s-1}}=\mathcal O(R^2),&\label{dGI11}\\[6pt]&\D\mathcal J^{~\,\m_1\dots\m_{s-1}}
_{1,\,\a_1\dots\a_{s-1}}=\D\tilde{\mathcal J}^{~\,\m_1\dots\m_{s-1}}_{1,\,\a_1\dots\a_{s-1}}+\mathcal O(R^2),&\label{dGI12}\\[6pt]
&\D\mathcal K^{~\,\m_1\dots\m_{s-2}}_{1,\,\a_1\dots\a_{s-1}}=\D\tilde{\mathcal K}^{~\,\m_1\dots\m_{s-2}}_{1,\,\a_1\dots\a_{s-1}}
+\mathcal O(R^2).&\eea{dGI13}
On the other hand, corresponding to $i=1$ and $i=2$ respectively, the second and third gauge identities call for
\beq \D \mathcal K^{~\,\m_1\dots\m_{s-2}}_{2,\,\a_1\dots\a_{s-2}}=-\a R\,\d^{\m_1\dots\m_{s-2}}_{\a_1\dots\a_{s-2}}+\mathcal O(R^2),\eeq{dGI2}
with all other corrections being only $\mathcal O(R^2)$.

\end{appendix}

\end{document}